\documentclass[aps,prl,twocolumn,showpacs,floatfix,superscriptaddress,amsmath,amssymb]{revtex4-2}
\usepackage[pdfusetitle]{hyperref}
\usepackage[dvipsnames]{xcolor}
\usepackage[final]{graphicx}
\usepackage{bm}
\usepackage{enumerate, enumitem}
\usepackage[normalem]{ulem}
\usepackage[capitalize]{cleveref}
\crefname{equation}{Eq.}{}
\crefrangeformat{equation}{Eqs.~(#3#1#4)-(#5#2#6)}

\graphicspath{{figs/paper1/}}

\newcommand{\postime}{(\bm{r},t)}

\newcommand{\Rl}{R_{\infty}}
\newcommand{\kl}{k_{\infty}}
\newcommand{\alphac}{\alpha_{c}}
\newcommand{\alphax}{\alpha_{\times}}

\newcommand{\movie}[1]{\protect {(see Movie #1)}}

\begin{document}

\title{Defect Solutions of the Non-reciprocal Cahn-Hilliard Model: Spirals and Targets}
\author{Navdeep Rana}
\affiliation{Max Planck Institute for Dynamics and Self-Organization (MPI-DS), D-37077 G\"ottingen, Germany}

\author{Ramin Golestanian}
\affiliation{Max Planck Institute for Dynamics and Self-Organization (MPI-DS), D-37077 G\"ottingen, Germany}
\affiliation{Rudolf Peierls Centre for Theoretical Physics, University of Oxford, Oxford OX1 3PU, United Kingdom}

\begin{abstract}
We study the defect solutions of the Non-reciprocal Cahn-Hilliard model (NRCH). We find two kinds of defects, spirals with unit magnitude topological charge, and topologically neutral targets. These defects generate radially outward travelling waves and thus break the parity and time-reversal symmetry. For a given strength of non-reciprocity, spirals and targets with unique asymptotic wavenumber and amplitude are selected. We use large-scale simulations to show that at low non-reciprocity $\alpha$, disordered states evolve into quasi-stationary spiral networks. With increasing $\alpha$, we observe networks composed primarily of targets. Beyond a critical threshold $\alphac$, a disorder-order transition from defect networks to travelling waves emerges. The transition is marked by a sharp rise in the global polar order.
\end{abstract}

\maketitle

\emph{Introduction.}---Constituents of active matter, biological or synthetic, interact in complex ways \cite{gompper2020}. These interactions are realized through various mechanisms, for example, chemical activity in colloids and enzymes \cite{soto2014, saha2019}, wake-mediated interactions in complex binary plasmas \cite{ivlev2015}, visual perception in bird flocks \cite{ballerini2008}, social communication in crowds of humans \cite{helbing1995, helbing2000, rio2018} and microswimmers \cite{dinelli2022}, tensorial hydrodynamic interactions in active carpets \cite{uchida2010}, and programmable logic in robots \cite{fruchart2021}. Breaking the action-reaction symmetry leads to novel features that are absent in equilibrium \cite{ivlev2015,loos2020}, including the possibility to engineer multifarious self-organization of building clocks in a choreographed manner \cite{osat2023}. Individuals in a chemically active mixture can assemble into self-propelling small molecules \cite{soto2014,soto2015} or form large comet-like clusters \cite{cohen2014,agudo-canalejo2019}. Non-reciprocal alignment interactions lead to a buckling instability of the ordered state in polar flocks \cite{dadhichi2020}, as well as a wide range of other novel features \cite{gupta2022,dekarmakar2022,knezevic2022}. In the recently introduced non-reciprocal Cahn-Hilliard model (NRCH) \cite{saha2020,you2020,saha2022}, parity and time-reversal (PT) symmetries break spontaneously which leads to the formation of travelling density bands \cite{saha2020, you2020}, coarsening arrest \cite{frohoff-hulsmann2021, saha2020}, and localized states \cite{frohoff-hulsmann2021a}. A variant of the NRCH model
with nonlinear non-reciprocal interactions exhibits chaotic steady states where PT symmetry is restored locally in fluctuating domains \cite{saha2022}. Although the NRCH model was introduced phenomenologically \cite{saha2020, you2020}, it has been highlighted recently that it is possible to derive it as a universal amplitude equation that emerge from a conserved-Hopf instability, occurring in systems with two conservation laws \cite{frohoff-hulsmann2023}. Moreover, it has been derived using systematic coarse-graining of a microscopic model of phoretic active colloids \cite{tucci2024}. 

\begin{figure}[b]
    \centering
    \includegraphics[width=\linewidth]{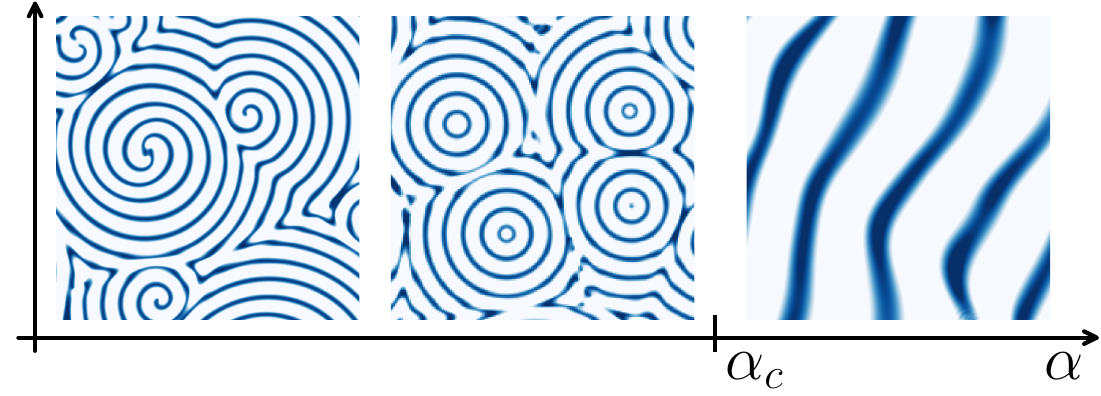}
    \caption{\label{fig:phase}
        Qualititative phase portrait for the NRCH model \eqref{eq:nrch} in the $\alpha$ space. A critical threshold $\alphac$ marks the onset of a disorder-order transition. When non-reciprocal interactions are weak ($\alpha \ll \alphac$), we find defect networks with isolated and bound spirals. With increasing $\alpha$, targets begin  to emerge and are the dominant defects right below $\alphac$. Above $\alphac$, noisy global polar order sets in. Fluctuations decay with time which eventually leads to travelling bands.
    }
\end{figure}

\begin{figure}
    \centering{\includegraphics[width=\linewidth]{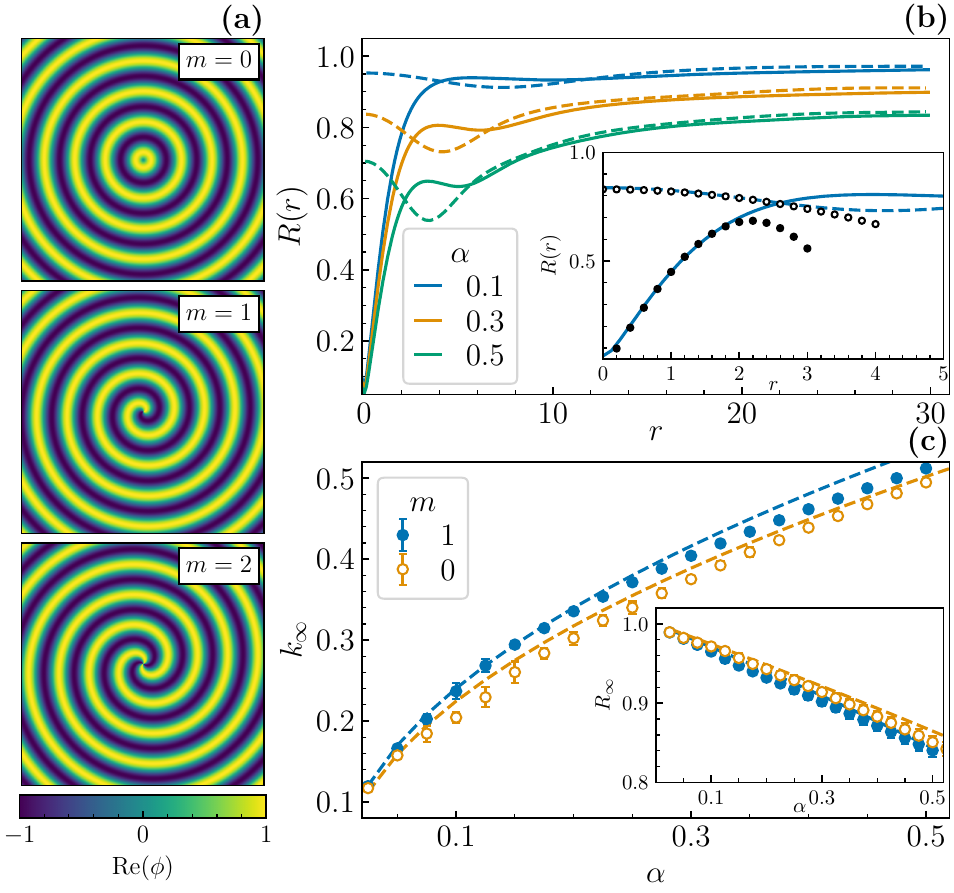}}
    \caption{\label{fig:defects} (a) Defect solutions of the NRCH model. (b) $R(r)$ vs. $r$ for $m=0~\rm{and}~1$ at different $\alpha \lesssim \alphax$. Inset: Comparison of $R(r)$ with small $r$ approximations \cite{Note1}. (c) Selected wave number $\kl$ vs. $\alpha$. We find $\kl = C \sqrt{\alpha}$ (dashed lines with same colours), where $C \sim 0.76$ for $m=1$ and $C \sim 0.7$ for $m=0$. Inset: $\Rl$ vs. $\alpha$ (dashed lines show $\sqrt{1-\kl^2}$). }
\end{figure}

Here, we study the defect solutions of the NRCH model \cite{saha2020, you2020}. We find two types of defects, spirals with a unit magnitude topological charge and topologically neutral targets (see \cref{fig:phase} and \cref{fig:defects}). They are the generators of travelling waves and thus break the PT symmetry. In addition, spirals break the chiral symmetry. Spirals are frequently observed and extensively studied in various systems described by the complex Ginzburg-Landau (CGL) equation, for example, the well-known Belousov-Zhabotinskii reaction, and colonies of \emph{Dictyostelium} \cite{hagan1982,zaikin1970, winfree1972, kuramoto1984, lee1996}. Topologically neutral targets are unstable in the framework of the CGL equation but can be stabilized by introducing spatial inhomogeneities
\cite{hagan1981, hendrey2000, aranson2002, kuramoto1984}. In the context of non-reciprocal interactions, creation and annihilation of spiral defects has been reported in the context of active turbulence in wet polar active carpets \cite{uchida2010}. Programmable robots are shown to break the chiral symmetry and spontaneously rotate in clockwise or anticlockwise manner \cite{fruchart2021}. 

\begin{figure*}
    \centering{\includegraphics[width=\linewidth]{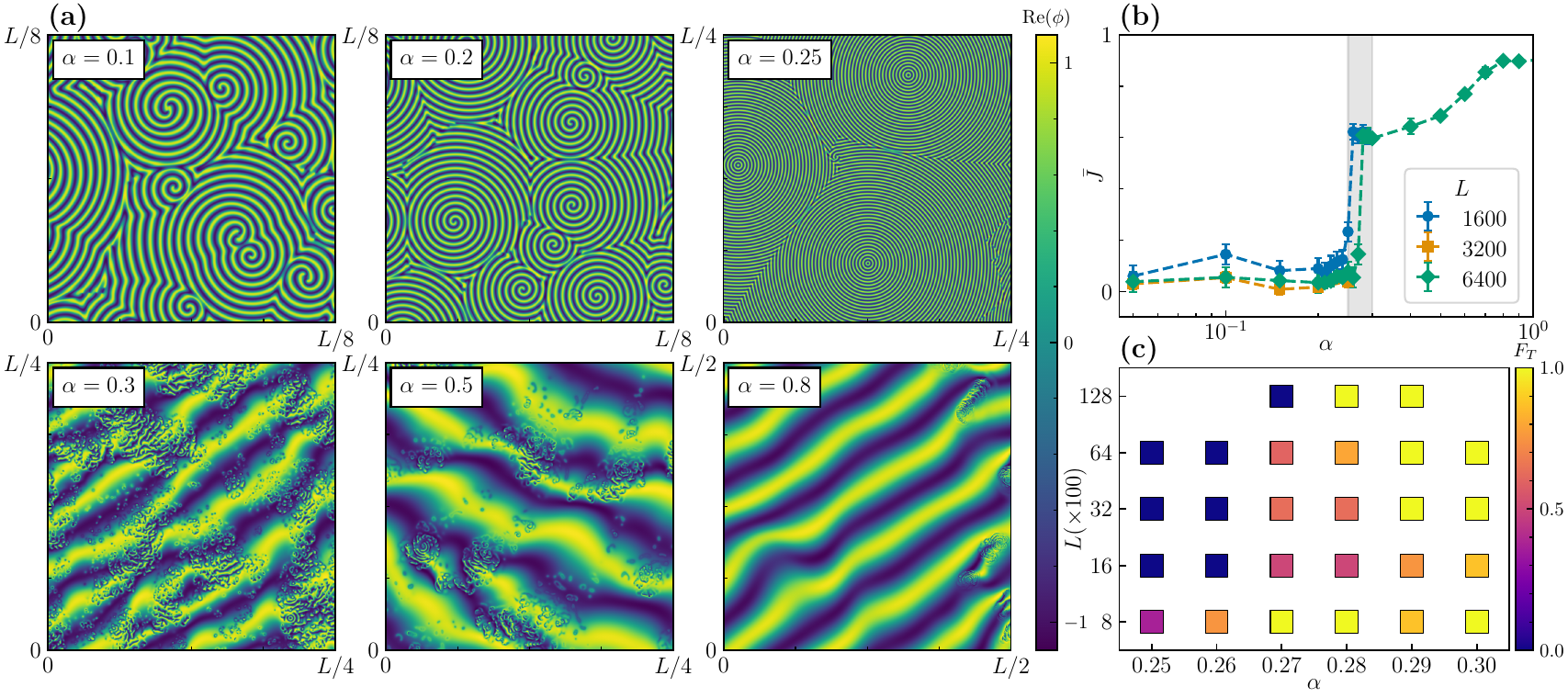}}
    \caption{\label{fig:transition}
        (a) Transition from disordered defect networks to travelling waves on increasing the non-reciprocity strength $\alpha$ for $L=6400$ and $N=4096$\cite{Note1}.
        We plot the real component of $\phi$ at long times on smaller subdomains for a better visual representation. (b) Global polar order in the steady state with varying $\alpha$. At $\alphac$, we observe a sharp order-disorder transition. (c) Plot of the fraction of simulations $(F_T)$ that transition to travelling waves for different $\alpha$ (shaded region in (b)) and $L$. For $\alpha \geq 0.28$, almost all of the simulations show the transition and hence we infer $\alphac = 0.28 \pm 0.01$. Smaller boxes can show transition at $\alpha \lesssim \alphac$ due to finite size effects \cite{Note1}.
    }
\end{figure*}

\begin{figure}
    \centering{\includegraphics[width=\linewidth]{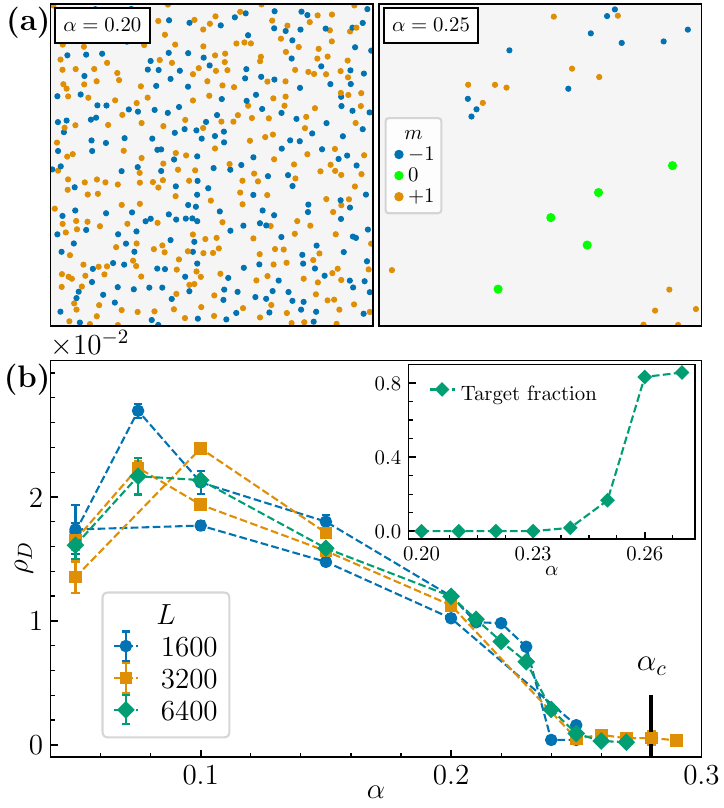}}
    \caption{\label{fig:density} (a) Scatter plot of spiral and target cores for two different values of $\alpha$ over the entire simulation domain $L=6400$. Targets show strikingly higher separation compared to spirals. (b) Defect density $\rho_D$, defined as the number of spirals and targets per unit area, decreases with increasing $\alpha$, in the vicinity of $\alphac$, $\rho_D \sim \mathcal{O}(L^{-2})$ and it vanishes above $\alphac$. Further, it is sensitive to the initial conditions, as shown by different lines for the same value of $L$. Inset: Target fraction increases with $\alpha$ $(L=6400)$.}
\end{figure}

\emph{Summary of results.}---Our central finding is that the NRCH model admits stable spiral and target defect solutions. Remarkably, no additional spatial inhomogeneities are needed to stabilize the targets \cite{hagan1981,hendrey2000}. For a given strength of non-reciprocal interactions $(\alpha)$, defect solutions with a unique asymptotic wave number $\left(\kl =
C\sqrt{\alpha}\right)$ and amplitude $\left(\Rl = \sqrt{1-\kl^2}\right)$ are selected (see \cref{fig:defects}). As a consequence of the wave number selection, defect solutions cease to exist beyond a crossover point $\alphax = 1/C^2$. However, in our large-scale numerical simulations starting from disordered states, defect solutions vanish for $\alpha$ well below $\alphax$ and we find a disorder-order transition at $\alphac \ll \alphax$ (see \cref{fig:phase} and \cref{fig:transition}). Below $\alphac$, an initially disordered state evolves into quasi-stationary defect network with no global polar order \movie{2}. While both kinds of defect are stable for a given $\alpha$, defect networks exhibit a clear preference for spirals or targets. At small $\alpha$, we exclusively find spirals. As we increase
$\alpha$, targets start to appear as well, and close to the transition point $\alpha \lesssim \alphac$, we find target-dominated defect networks (see \cref{fig:transition,fig:density}). Above $\alphac$, we find travelling waves that show global polar order, rendered imperfect by mesoscopic fluctuations that decay with time and eventually lead to travelling bands. A sharp jump in the global polar order marks the onset of this transition (see \cref{fig:transition}).

\emph{Model.}---We consider a minimal model of two conserved scalar fields $\phi_{1}\postime$ and $\phi_{2}\postime$ with non-reciprocal interactions. The complex scalar order parameter $\phi = \phi_1 + i \phi_2$ obeys the following non-dimensional equation \footnote{Supplemental Material available at ????. It includes a discussion on the NRCH model and its various solutions, large and small $r$ behaviour of $R(r)$ and $k(r)$, details of the numerical methods, finite-size effects on the defect solutions, and description of the movies.}
\begin{equation}\label{eq:nrch}\begin{aligned}
    \partial_t{\phi} &= \nabla^2\left[(-1 + i \alpha)\phi + |\phi|^2\phi - \nabla^2 \phi\right],
\end{aligned}\end{equation}
where the parameter $\alpha$ quantifies the non-reciprocal interactions and $\alpha>0$ implies that $\phi_1$ chases $\phi_2$. Conservation of particle numbers for both species makes it impossible to eliminate $\alpha$ using a global phase transformation as is customarily done for the CGL equation \cite{aranson2002}. The length scales of interest are the system size $L$, the spinodal instability cutoff length $\ell$ which for \eqref{eq:nrch} is set to unity, and the length scale $\ell_\alpha = \ell/\sqrt{\alpha}$ that governs the oscillatory features of \eqref{eq:nrch} \cite{Note1}. Travelling wave solutions of \eqref{eq:nrch} have the form
\begin{equation}\label{eq:wave}\begin{aligned}
    \phi\postime = R \, e^{i\left(\bm{k}\cdot \bm{r} - \omega t\right)},
\end{aligned}\end{equation}
with $k=|\bm{k}|$, $R=\sqrt{1-k^2}$, and $\omega = \alpha k^2$.
For any $\alpha$, an infinite number of plane waves with $k < 1$ are possible. Waves with $k < 1/\sqrt{3}$ are linearly stable and small perturbations at wave number $q$ decay with a rate $ \propto
\mathcal{O}(q^2)$, while beyond this threshold the Eckhaus instability sets in \cite{zimmermann1997,aranson2002,saha2022}.

\emph{Defect solutions.}---We now show that the NRCH model \eqref{eq:nrch} admits defect solutions of the form
\begin{equation}\label{eq:defect}\begin{aligned}
    \phi(\bm{r},t) = R(r) \, e^{i\left[ m \theta + Z(r) - \omega t \right]},
\end{aligned}\end{equation}
where $r$ and $\theta$ represent the polar coordinates, and $r$ is measured from the defect core. $R(r)$ is the
amplitude, $Z(r)$ is the phase, and $m$ is the topological charge. Figure \ref{fig:defects}(a) shows defect solutions for different $m$. Topologically neutral target $(m=0)$ and charged spiral $(m=\pm 1)$ are stable, whereas defects with $|m|>1$ are not and they evolve into bound pair of spirals.
In \cref{fig:defects}(b), we plot $R(r)$ vs. $r$ for various values of $\alpha$ and $m$. An isolated spiral core is
singular and stationary, thus $R(r)$ vanishes at the origin and is independent of time. On the other hand, $R(r)$ is
finite at the core of a topologically neutral target and oscillates slowly with time \cite{hagan1982, aranson2002,
hendrey2000}. At small $r$, we find $R(r) \sim a_1 r - a_3 r^3$ for spirals and $R(r) \sim a_0 - a_2 r^2$ for targets. For both spiral and targets, we obtain $k(r) \equiv \frac{d Z}{dr} \sim b_1 r - b_3 r^3$ at small $r$ \cite{Note1}.

Defects are the generators of plane waves that propagate outwards in the radial direction \movie{1}. Thus, at large distances from the defect core $(r \gg 1)$, the wave front approaches that of a plane wave, i.e. $k(r) \rightarrow \kl$, and $R(r) \rightarrow \Rl = \sqrt{1 - \kl^2}$. We require $\omega = \alpha \kl^2$ to ensure proper oscillations at all $r$ and $\kl > 0$ to have a radially outward group velocity i.e. $v_g = 2\alpha \kl > 0$ as $\alpha > 0$. To first order in $1/r$, $R(r) \sim \Rl + \frac{1}{r}\frac{\alpha^2+\kl^4}{2 \alpha  \kl \Rl}$ and $k(r) \sim \kl + \frac{1}{r} \frac{\kl^2}{2 \alpha }$ \cite{Note1}. Emitted plain waves screen the defect core from outside perturbations \cite{aranson1993} and their stability implies that the defect solutions are also stable at large distances from their core. In \cref{fig:defects}(c), we show that $\kl$ increases with $\alpha$ and using a numerical fit we find that $ \kl = C \sqrt{\alpha}$; the inset verifies the relation between $\Rl$ and $\kl$. It is evident that $\Rl$ vanishes for $\alpha \geq \alphax \equiv 1/C^2$ and the defect solutions cease to exist beyond $\alphax$. Stability of the plane waves then implies a potential crossover from defects to travelling waves for $\alpha \geq \alphax$. The Eckhaus instability further reduces $\alphax$ as the far field wave fronts are unstable for $\kl > 1/\sqrt{3}$ and we obtain $\alphax = 1/3C^2 \sim 0.58$.

\emph{Defect networks in simulations.}---At equilibrium $(\alpha = 0)$, a system quenched from a high temperature disordered state to sub-critical temperatures undergoes bulk phase separation. The phase separated domains grow with time and a unique growing length scale characterizes this coarsening dynamics \cite{bray2002, onuki2002}. The dense-dense (liquid-liquid) or dense-dilute (liquid-gas) coexistence states are determined by the reciprocal interactions between the two scalar fields. For $\alpha \neq 0$, the interplay of non-reciprocity with equilibrium forces allows for the emergence of \emph{a priori} undetermined complex spatio-temporal patterns; a hallmark feature of non-equilibrium systems \cite{cross1993, marchetti2013, ramaswamy2010, uchida2010}. To understand the dynamical features of the NRCH model, we perform large scale simulations of \eqref{eq:nrch} with varying non-reciprocity $\alpha$ and system size $L$ \cite{Note1}.
In \cref{fig:transition}(a), we show the typical steady-state solutions obtained from the evolution of a disordered state. A critical threshold of non-reciprocal interactions, $\alphac < \alphax$, separates the phase space into two distinct regimes -- quasi-stationary defect networks and travelling waves superimposed with local fluctuations \cite{Note1}. In what follows, we highlight the main features of these non-equilibrium states; a detailed analysis will be presented elsewhere. 

For $\alpha < \alpha_c$ [see \cref{fig:transition}(a) (top panels) and \cref{fig:density}(a)], after an initial
transient period in which numerous newly-born defects move around and merge, a quasi-stationary steady-state emerges
\movie{2}. This steady configuration is sensitive to the initial conditions and shows limited dynamics. The arms of the
isolated spirals rotate, the targets pulsate, and bound pairs of like-charged spirals orbit around a common centre.
Along the disinclination lines, where the waves emitted from spiral or target cores meet, we find small clusters of
additional defects that cannot be classified as spirals or targets. These defects show dynamical rearrangement with time
and form locally unsteady patches in an otherwise frozen network \movie{2}. The composition of defect networks changes
with $\alpha$. For small $\alpha$, we observe isolated spirals and a few bound pairs of like-charged spirals that orbit
around a common centre. As we increase $\alpha$, targets emerge as well and right below $\alphac$, we primarily find
targets. For a given $\alpha$, targets show a strikingly higher inter-defect separation as compared to
the spirals [see \cref{fig:density}(a, b)]. Due to the sensitivity to the initial conditions, we observe variability
in the defect density, but the overall behaviour exhibits an initial increase that is then followed by a decrease, as $\alpha$ is increased [see \cref{fig:density}(b)].

For $\alpha \gtrsim \alphac$, we find travelling waves [see \cref{fig:transition}(a) (bottom panels)]. The transient
period shows a mixture of defects and growing patches of polar order, which quickly washes away the defect cores.
Afterwards, global polar order, albeit marred by local spatial fluctuations, emerges. These fluctuations decay with time
and eventually we find travelling density bands. To quantify the transition from defect networks to travelling waves we
compute the average polar order $\bar{J} \equiv \left|\left<\bm{\widehat{J}}\postime\right>\right|$, where
$\bm{J}\postime \equiv \frac{1}{2 i}\left(\phi^{*} \nabla \phi - \phi\nabla\phi^{*}\right) = \phi_1\nabla\phi_2 -
\phi_2\nabla\phi_1$ is the polar order parameter and $\left<\ldots\right>$ implies averaging over space and time in the
steady-state. For a monochromatic plane wave of the form \eqref{eq:wave}, $\bm{J}=R^2\bm{q}$, and thus $\bar{J} = 1$. On
the other hand, for defect solutions \eqref{eq:defect}, we have $\bm{J} = R(r)^2 \left(k(r) \hat{r} + \frac{m}{r}
\hat{\theta}\right)$, which implies $\bar{J} \sim 0$. Far away from the defect cores $(r \gg \ell)$, we obtain $\bm{J}
\sim \Rl^2 \kl \hat{r}$, which is independent of the value of $m$. The defects emanate radially outward travelling waves
and $\bm{J}$ has a topological singularity with unit positive charge at the defect core.

As shown in \cref{fig:transition}(b), the transition from defect networks to travelling waves is marked by a sharp increase in the average polar order $\bar{J}$ at $\alpha = \alphac$. We find that $\bar{J} \sim 0$ for defect networks $(\alpha < \alphac)$, while it acquires a finite value for the travelling wave states $\alpha > \alphac$. Spatial fluctuations decrease with increasing $\alpha$, thus $\bar{J} < 1$ for $\alpha \gtrsim \alphac$, while it saturates close to its maximum permissible value
$\bar{J}=1$ for $\alpha \gg \alphac$. From \cref{fig:transition}(b), we expect $0.25 < \alphac < 0.30$, however, in the vicinity of $\alphac$, finite size effects can influence the steady states, especially for smaller box sizes. To obtain a better estimate of $\alphac$, we have performed an ensemble of numerical simulations with varying box sizes $(L)$ spanning two orders of magnitude. In \cref{fig:transition}(c), we show the fraction of simulations that reached a travelling wave steady state for various $L$ and $\alpha$. From these simulations, we infer $\alphac \sim 0.28 < \alphax$.

\emph{Discussion.}---Non-reciprocal interactions emerge naturally in non-equilibrium systems with complex interactions \cite{golestanian2022,soto2014,ivlev2015}, and this effective breaking of the action-reaction symmetry leads to a variety of novel features. Here, we have unveiled a new feature of non-reciprocal interactions, namely, the emergence of topological defects in binary mixtures of conserved scalar densities. We find two kind of defects for the NRCH model, spirals with a unit magnitude topological charge and neutral targets. For a given $\alpha$, defects with a unique wavenumber are selected, which immediately predicts a crossover from defects to plain waves at $\alpha = \alphax$. However, our large-scale numerical simulations show a disorder-order transition from quasi-stationary defect networks to imperfect global polar order $\alpha = \alphac < \alphax$. These states show a rich phase space behaviour. While both charged and neutral defects are allowed for any $\alpha < \alphax$, at low values of $\alpha$ the system prefers topologically charged disorder, where the chiral symmetry is broken, and we find isolated and bound pairs of spirals. Close to $\alphac$, disorder is topologically neutral and targets are the preferred defects. Above $\alphac$, noisy travelling waves with spontaneously broken polar symmetry emerge. The fluctuations in these states decay with time, but can persist for a very long duration especially for $\alpha \gtrsim \alphac$.

Our study uncovers important features concerning the phenomenology of active matter with non-reciprocal interactions. We note here that while the isolated defect solutions are stable in the presence of persistent noise, wave interaction and finite size effects can result in interesting pattern formation \movie{3}. A natural next step will be to study the stability of isolated defect solutions to small perturbations, as has been in the case of the CGL equation \cite{hagan1982,aranson2002}, for which it has been observed that the defect network states are not static but evolve extremely slowly \cite{brito2003}. It will be interesting to investigate if the defect networks in the NRCH model exhibit the phases of vortex liquid and vortex glass with intermittent slow relaxation observed in CGL equation. We have focused here on a simplified version of the NRCH model with purely non-reciprocal interactions at the linear level and restored global rotational symmetry in the $\phi-$space \cite{Note1}. In the Supplemental Material \cite{Note1} we show the defect solutions in the presence of linear reciprocal interactions. In the future, our study could be extended to the study of the properties of these defect solutions and to include nonlinear non-reciprocal interactions \cite{saha2020}. Finally, in our studies here we have neglected the naturally occurring noise in the NRCH equation, which is conserved and has been found not to affect the stability of long-range polar order in the travelling-bands phase \cite{pisegna2024}. We note that the existence of noise can help introduce defects in the layered phase, which may (or may not) destabilize the smectic order in the system, as examined in a number of related studies \cite{toner1981,chen2013,julicher2022}. Such considerations will be relegated to future work.

\begin{acknowledgements}
We thank Giulia Pisegna and Suropriya Saha for fruitful discussions. We acknowledge support from the Max Planck School Matter to Life and the MaxSynBio Consortium which are jointly funded by the Federal Ministry of Education and Research (BMBF) of Germany and the Max Planck Society.
\end{acknowledgements}
\bibliography{Bibliography}
\bibliographystyle{apsrev4-2}

\end{document}


\title{Defect Solutions of the Non-reciprocal Cahn-Hilliard Model: Spirals and Targets \\
\textit{Supplemental Material}}
\author{Navdeep Rana}
\affiliation{Max Planck Institute for Dynamics and Self-Organization (MPI-DS), D-37077 G\"ottingen, Germany}

\author{Ramin Golestanian}
\affiliation{Max Planck Institute for Dynamics and Self-Organization (MPI-DS), D-37077 G\"ottingen, Germany}
\affiliation{Rudolf Peierls Centre for Theoretical Physics, University of Oxford, Oxford OX1 3PU, United Kingdom}

\date{\today}

\maketitle

\tableofcontents

\setcounter{equation}{0}
\setcounter{figure}{0}
\renewcommand{\theequation}{S\arabic{equation}}
\renewcommand{\thefigure}{S\arabic{figure}}

\newpage

\section{\label{sec:model} The Non-reciprocal Cahn-Hilliard model}

We consider a two species non-reciprocal Cahn-Hilliard (NRCH) model of phase-separating mixtures \cite{saha2020,you2020,saha2022}. The system is described by the conserved scalar fields $\phi_1\postime$ and $\phi_2\postime$. The equations of motion for $\phi_{1,2}$ are
\begin{equation}\label{eq:nrch-full}\begin{aligned}
    \partial_t{\phi_1} &= \nabla^2\left[-a\phi_1 + b\phi_1^3 + \left(\chi - \alpha\right) \phi_2 + 2\chi'\phi_1 \phi_2^2 - K \nabla^2 \phi_1\right], \\
    \partial_t{\phi_2} &= \nabla^2\left[-a\phi_2 + b\phi_2^3 + \left(\chi + \alpha\right) \phi_1 + 2\chi'\phi_2 \phi_1^2 - K \nabla^2 \phi_2\right],
\end{aligned}\end{equation}
where $a, b$, and $K$ fix the form of the Ginzburg-Landau free energy, and $\chi$ and $\chi'$ encode the equilibrium interactions between the two species. Non-reciprocal interactions at the linear level are governed by the single parameter $\alpha$. For $\alpha=0$, \eqref{eq:nrch-full} reduces to a two species binary mixture, which shows bulk phase separation with dense-dense (liquid-liquid) or dense-dilute (liquid-gas) coexistence depending on the values of $\chi$ and $\chi'$. $\alpha > 0$ implies that $\phi_1$ chases $\phi_2$.

Non-reciprocity has far-reaching consequences on the dynamics of the system. Parity and time-reversal symmetries no longer hold and travelling density bands are observed instead of bulk phase separation. Though travelling states are generic features of the full NRCH model \eqref{eq:nrch-full}, it is easier to obtain analytical dispersion relations for the values $\left(\chi=0, \chi'=\frac{1}{2}\right)$, where the reciprocal interactions are chosen in a way to leave the model invariant under a constant phase change in $\phi$ \cite{saha2022}. We begin by recasting \eqref{eq:nrch-full} with the transformation
\[
\phi \equiv \sqrt{\frac{a}{b}} \left(\phi_1 + i \phi_2\right),\]
and choose $K=1$, $a=1$, which sets the length scale $\ell=\sqrt{K/a}$ and the time scale $\tau = K/a^2$ to unity. We obtain the following equation with a single control parameter $\alpha$
\begin{equation}\label{eq:nrch-supp}\begin{aligned}
    \partial_t{\phi} &= \nabla^2\left[(-1 + i \alpha)\phi + |\phi|^2\phi - \nabla^2 \phi\right].
\end{aligned}\end{equation}
$\ell$ is the length scale above which perturbations to the $\phi=0$ state are unstable and $\tau$ determines the growth rate of these perturbations. The Fourier amplitude of the perturbation $\delta\phi = \widehat{\phi}(\bm{k}, t) e^{i\bm{k}\cdot\bm{x}}$ evolves as
\begin{equation}\begin{aligned}
    \partial_t \widehat{\phi}(\bm{k}, t) = k^2(1-i\alpha - k^2) \widehat{\phi}(\bm{k}, t),
\end{aligned}\end{equation}
implying an oscillatory instability $\forall~ 0 < k < \ell^{-1}$, with a growth rate $k^2(1 - k^2)$ and oscillation
frequency $\alpha q^2$.

Note that rescaling space and time as $~\bm{r} \rightarrow \ell \bm{r}',~\mathrm{and}~t \rightarrow \tau t'$ in \eqref{eq:nrch-full} will also lead to \eqref{eq:nrch-supp}. The oscillatory features of \eqref{eq:nrch-full} are governed by the length and time scales $\ell_\alpha\equiv\sqrt{\frac{K}{\alpha}}$ and $\tau_\alpha = \frac{K}{\alpha^2}$ respectively. Equations \eqref{eq:nrch-full} and \eqref{eq:nrch-supp} are invariant under the simultaneous transformations $\left(\phi_1, \phi_2, \alpha\right) \rightarrow (\phi_2, \phi_1, -\alpha)$, thus we can always assume $\alpha > 0$. Additionally, the transformation $\phi\postime \rightarrow e^{i \Phi} \phi\postime$, where $\Phi$ is a constant phase, also leaves \eqref{eq:nrch-supp} unchanged.

The plain wave ansatz $\phi\postime = R \, e^{i(\bm{k}\cdot\bm{r} -\omega t)}$ is a solution of \eqref{eq:nrch-supp} with $$\omega =\alpha k^2, ~ R = \sqrt{1 - k^2}.$$ In other words, \eqref{eq:nrch-supp} admits a plain wave solution with $k < 1$ whose amplitude and frequency are wave number $(k)$ dependent. As mentioned in the main text, waves with $k < 1/\sqrt{3}$ are linearly stable and small perturbations at wave number $q$ decay with a rate $ \propto
\mathcal{O}(q^2)$ \cite{zimmermann1997,aranson2002,saha2022}.

\section{\label{sec:defects} Defect solutions}

In this section, we present results for the defect solutions of the NRCH model. The defect solutions of
\eqref{eq:nrch-supp} are of the form
\begin{equation}\label{eq:defect-ansatz}\begin{aligned}
    \phi(\bm{r},t) = R(r) \, e^{i\left( m \theta+Z(r) - \omega t\right)},
\end{aligned}\end{equation}
where $r$ and $\theta$ represent the polar coordinates, and $r$ is measured from the defect core. $R(r)$ is the
amplitude, $Z(r)$ is the phase, and $m$ is the topological charge. Note that the defect solutions are invariant under
the transformation $\alpha \to -\alpha,~Z(r) \to -Z(r)$.

$R(r)$ and $k(r)$ can be obtained either by substituting the defect ansatz in \eqref{eq:nrch-supp} and solving the
resulting coupled equations or by directly simulating \eqref{eq:nrch-supp} with the ansatz \eqref{eq:defect-ansatz} as the
initial condition. We have used the latter method to obtain the data shown in main text. In what follows, we present some analytical results for $R(r)$ and $k(r)$.

\subsection{Equations for $R(r)$ and $k(r)$}
The Laplacian of the defect ansatz can be written as
\begin{equation}\begin{aligned}
    \nabla^2\phi(\bm{r},t) = \left(\FD{R(r)} + i \GD{R(r)}\right)e^{i \Phi(r,\theta,t)},
\end{aligned}\end{equation}
where we have defined
\begin{equation}\begin{aligned}
    \Phi(r,\theta,t) &= m \theta + Z(r) - \omega t, \\
    \FD{f(r)} &= f''(r) + \frac{f'(r)}{r} - \left(k(r)^{2} + \frac{m^2}{r^2}\right)f(r),\\
    \GD{f(r)} &= k'(r)f(r) + 2 k(r) f'(r) + \frac{k(r)f(r)}{r},
\end{aligned}\end{equation}
and $k(r) \equiv d Z/d r$. Substituting the defect ansatz in the NRCH equation \eqref{eq:nrch-supp}, we obtain
\begin{equation}\begin{aligned}
    i \omega R(r) e^{i \Phi(r,\theta,t)} &= \nabla^2\left\{\left[R(r) - R(r)^{3} + \FD{R(r)} - i \alpha R(r) + i\GD{R(r)}\right]e^{i \Phi(r,\theta,t)}\right\}.
\end{aligned}\end{equation}
Separating the real and imaginary parts, we arrive at the following equations for $R(r)$ and $k(r)$.
\begin{equation}\label{eq:frgr}\begin{aligned}
   & \FD{R(r) - R^3(r) + \FD{R(r)}} + \GD{\alpha R(r) - \GD{R(r)}} = 0, \\&\mathrm{and}~\\
   & \GD{R(r) - R^3(r) + \FD{R(r)}} - \FD{\alpha R(r) - \GD{R(r)}} - \omega R(r) = 0.
\end{aligned}\end{equation}
To ensure consistent oscillatory behaviour at all distances from the defect core requires $\omega = \alpha \kl^2$. With this, the full form of the above equations is $\sum_{n=0}^{4} A_n r^n = 0$ and $\sum_{n=0}^{3} B_n r^n = 0$, where the coefficients are
\begin{equation}\begin{aligned}
    A_4 &=  R_{rrrr}+R \left(-3 R R_{rr}+\alpha  k_{r}-3 k_{r}^2-6 R_{r}^2\right)+2 k \left(R_{r} \left(\alpha -6 k_{r}\right)-2 R k_{rr}\right),\\
        &~~~~+k^2 \left(-6 R_{rr}+R^3-R\right)+R_{rr}+k^4 R, \\
    A_3 &=  2 R_{rrr}+k R \left(\alpha -6 k_{r}\right)-3 \left(2 k^2+R^2\right) R_{r}+R_{r}, \\
    A_2 &=  -\left(2 m^2+1\right) R_{rr}+R \left(\left(2 m^2+1\right) k^2-m^2\right)+m^2 R^3, \\
    A_1 &=  \left(2 m^2+1\right) R_{r}, \\
    A_0 &=  m^2 \left(m^2-4\right) R, \\
    B_3 &=  R \left(k_{rrr}+k^2 \left(\alpha -6 k_{r}\right)+k_{r}-\alpha  \kl^2\right)+4 k R_{rrr}+2 \left(2 k_{rr}-2 k^3+k\right) R_{r}\\
        &~~~~-R_{rr} \left(\alpha -6 k_{r}\right)+R^3 \left(-k_{r}\right)-6 k R^2 R_{r}, \\
    B_2 &=  2 R k_{rr}+k \left(6 R_{rr}-R^3+R\right)-R_{r} \left(\alpha -6 k_{r}\right)-2 k^3 R, \\
    B_1 &=  R \left(\alpha  m^2-\left(2 m^2+1\right) k_{r}\right)-2 \left(2 m^2+1\right) k R_{r}, \\
    B_0 &=  \left(2 m^2+1\right) k R. \\
\end{aligned}\end{equation}

For brevity, we have used $R = R(r), ~ k = k(r), ~ R_r = \frac{d R}{d r},~R_{rr} = \frac{d^2 R}{d r^2}$ etc. in the
above expressions.

As discussed in the main text, the boundary conditions for $R(r)$ and $k(r)$ are determined by the kind of defect (value
of $m$) and the fact that at large $r$, the wave front approaches that of a plane wave. A spiral carries a topological
singularity at its core and thus $R(r)$ has to vanish at $r=0$. No such restrictions apply to a target and $R(r)$ does
not vanish at the core.

\begin{figure}
    \centering{\includegraphics[width=0.8\textwidth]{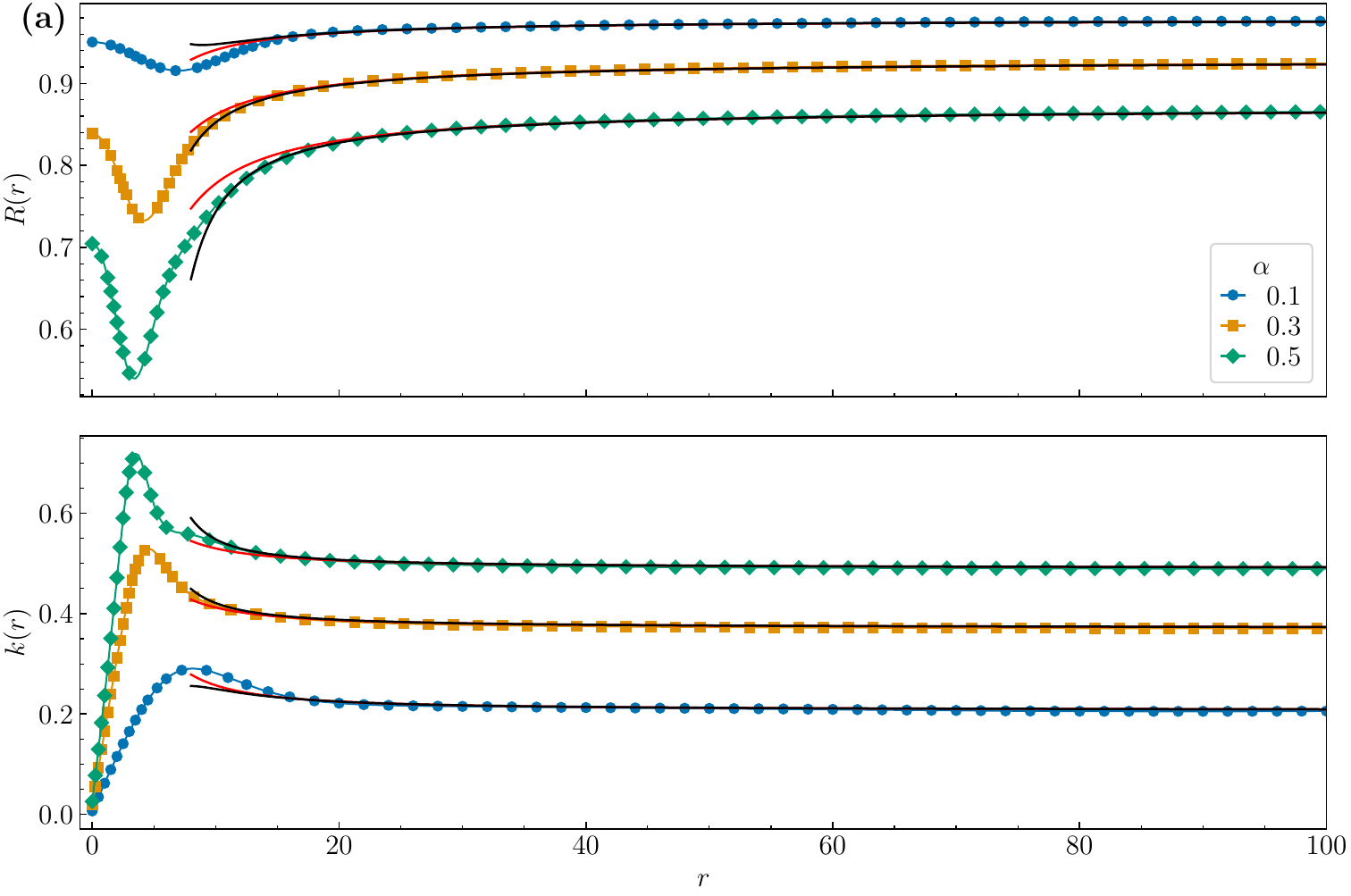}}
    \centering{\includegraphics[width=0.8\textwidth]{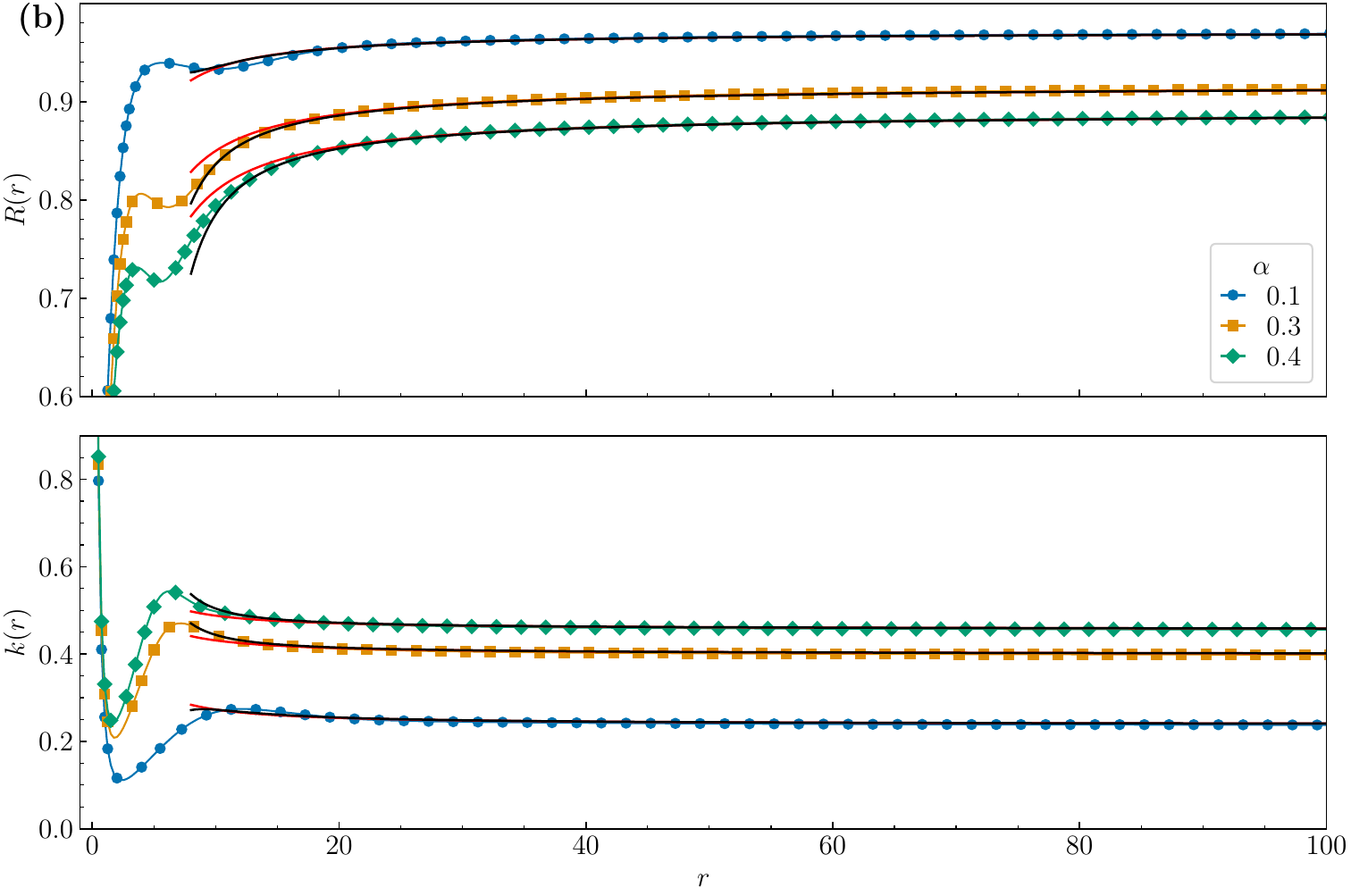}}
    \caption{ \label{fig:larger} Comparison of the numerical profiles of $R(r)$ and $k(r)$ for (a) targets and (b) spirals with the series expansion \eqref{eq:larger} and \eqref{eq:pnqn} where we have kept up to $n=3$ (in red) and $n=6$ (in black) terms and used $\kl$ obtained from the simulations.}
\end{figure}

\subsection{Large $r$ behavior}

Defects are the generators of plane waves travelling outwards in the radial direction. Thus, at large distances from the
defect core $(r \gg 1)$, the wave front approaches a plane wave, i.e. $k(r) \to \kl$, and $R(r) \to \Rl = \sqrt{1 -
\kl^2}$, as $r \to \infty$. To find the $r$ dependent corrections to $\Rl$ and $\kl$ for $r \gg 1$, we apply the
transformation $r \to \frac{1}{s}$ to \eqref{eq:frgr} and substitute the series ansatz
\begin{equation}\label{eq:larger}\begin{aligned}
    R(s) = \Rl\sum_{n=0}^{\infty}p_n s^n,~~k(s) = \kl\sum_{n=0}^{\infty}q_n s^n,
\end{aligned}\end{equation}
Equating the coefficients of all powers of $s$ to zero, we can iteratively solve for the coefficients $p_n$'s and $q_n$'s as functions of $\alpha$ and $\kl$. Below we list first four coefficients for both $R(r)$ and $k(r)$. As shown in \cref{fig:larger}, the large $r$ expressions match very well with the numerical profiles.
\begin{equation}\label{eq:pnqn}\begin{aligned}
    p_0 &=  1,
    ~p_1 =  \frac{\alpha^2+\kl^4}{2 \alpha  \kl \Rl^2},
    ~p_2 =  -\frac{\left(\alpha^2+\kl^4\right) \left(5 \alpha^2+\kl^4+4 \kl^2\right)}{4 \alpha^2 \kl^2 \Rl^4}, \\
    p_3 &=  -\frac{3 \left(\alpha^2+\kl^4\right) \left(-29 \alpha^4-2 \alpha^2 \left(2 m^2 \Rl^4+9 \kl^4+22 \kl^2-2\right)+\left(2 \kl^4-26 \kl^2-5\right) \kl^4\right)}{8 \alpha^3 \kl^3 \Rl^6}, \\
    q_0 &=  1,
    ~ q_1 =  \frac{\kl}{2 \alpha },
    ~ q_2 =  -\frac{4 \alpha^2 \left(m^2 \Rl^2+1\right)+\left(\kl^2+3\right) \kl^4}{4 \alpha^2 \kl^2 \Rl^2}, \\
    q_3 &=  \frac{3 \left(4 \alpha^2 m^2 \kl^2 \Rl^4-\kl^{10}+12 \kl^8+\kl^6+6 \alpha^4 \left(\kl^2+1\right)+4 \alpha^2 \left(\kl^2+5\right) \kl^4\right)}{4 \alpha^3 \kl^3 \Rl^4}.
\end{aligned}\end{equation}

\FloatBarrier

\begin{figure}
    \centering{\includegraphics[width=\textwidth]{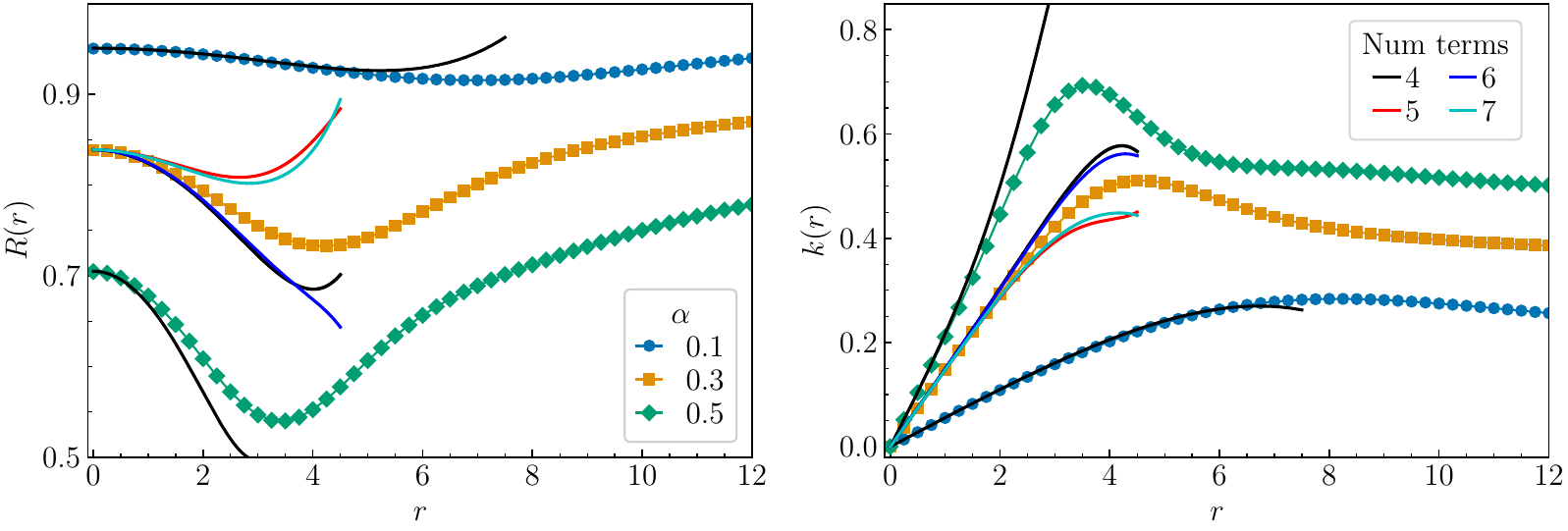}}
    \caption{ \label{fig:smallr-targets} Comparison of the small $r$ numerical profiles of $R(r)$ and $k(r)$ for targets with
        the series expansion \eqref{eq:smallr} where we have used numerical values of $R(0)=a_0$ and $k'(0)=b_1$. While
        the series solution matches well for small $\alpha$ with the numerical results, for larger $\alpha$ the series
        solution diverges rapidly. As we show for $\alpha=0.3$, keeping larger and larger number of terms in the series
    doesn't improve the match.}
\end{figure}

\subsection{Small $r$ behavior}

A direct substitution of the series ansatz
\begin{equation}\label{eq:smallr}\begin{aligned}
    R(r) = \sum_{n=0}^{\infty}a_n r^n,~~ k(r) = \sum_{n=0}^{\infty}b_n r^n,
\end{aligned}\end{equation}
in \eqref{eq:frgr} reveals that the $R(r)$ series for a spiral contains only odd terms, whereas for a target only the even terms survive. The $k(r)$ series for both spiral and target contains only odd terms. The boundary conditions at $r=0$ are
\begin{itemize}
    \item For a spiral, $R(0)=0$, $R'(0)=a_1$, $k(0)=0$, $k'(0)=b_1$.
    \item For a target, $R(0)=a_0$, $R'(0)=0$, $k(0)=0$, $k'(0)=b_1$.
\end{itemize}

Substituting a truncated small $r$ series expression in \eqref{eq:frgr} yields coupled nonlinear algebraic equations for
the series coefficients. It is possible to solve for the series coefficients provided the boundary conditions at $r=0$
are known, for example, from the numerical simulations. For targets, it is straightforward to obtain the numerical
values of $a_{0}$ and $b_{1}$. However, we find that the resulting series solutions diverge rapidly from the numerical
solutions for $r > 1$ (see \cref{fig:smallr-targets}) and are of little use to understand the behaviour of targets.

\begin{figure}
    \centering{\includegraphics[width=1.0\textwidth]{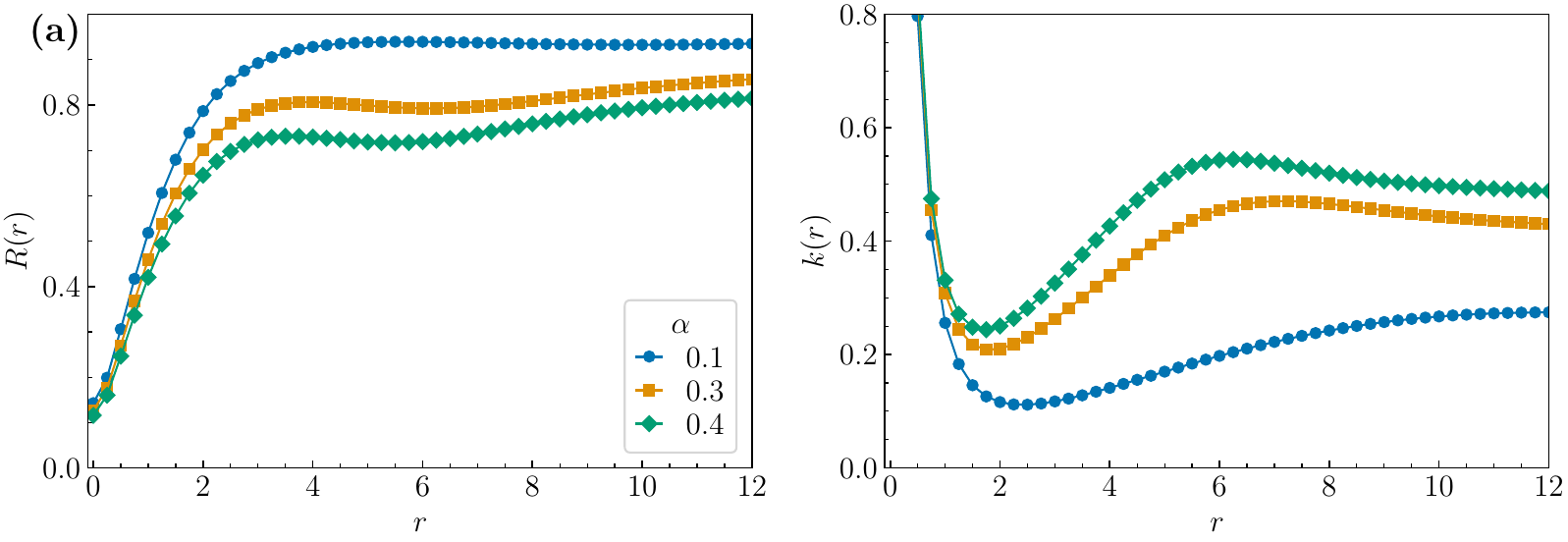}}
    \centering{\includegraphics[width=1.0\textwidth]{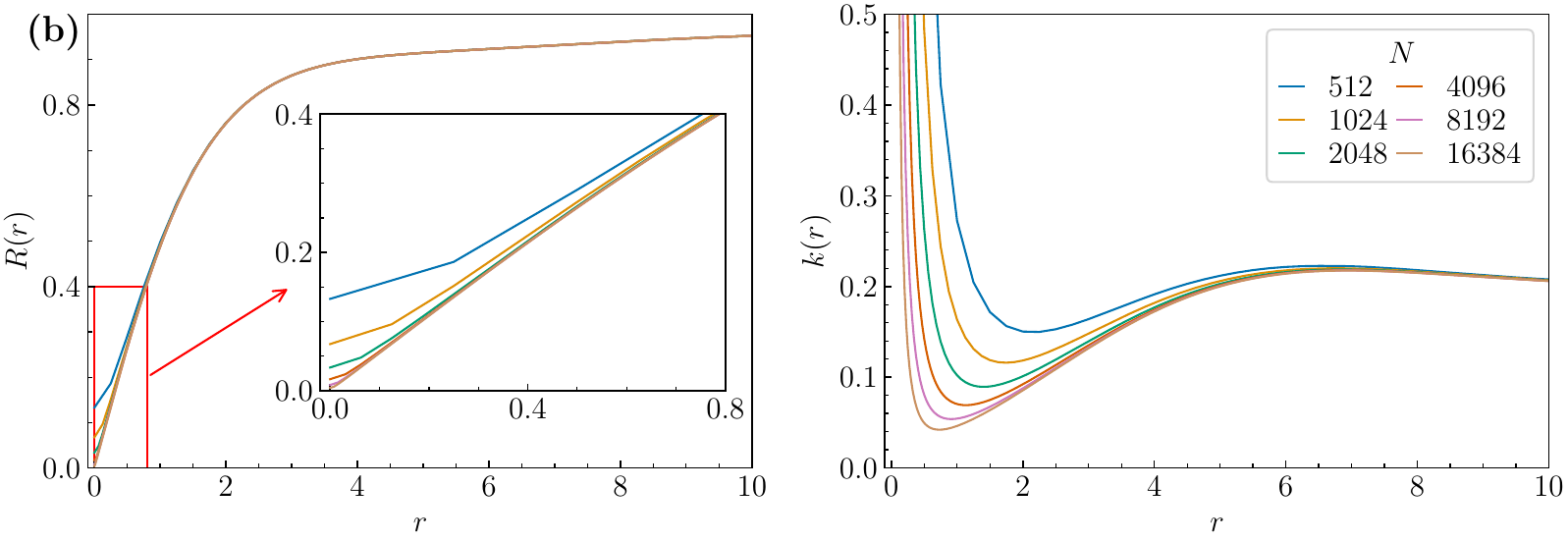}}
    \caption{\label{fig:smallr-spirals} (a) Small $r$ profiles for spirals at different $\alpha$. $R(r)$ never truely
        vanishes and  $k(r)$ diverges at $r=0$ because of the topological singularity and the fact that the cartesian
        grid used to simulate the NRCH model lacks the polar symmetry of the defect solutions. (b) Comparison of small
        $r$ profiles for fixed $\alpha=0.2$ and $L=128$ for different resolution $N$. $R(0)$ is closer and closer to
        it's true value and the singular region for $k(r)$ grows smaller and smaller with increasing $N$. Inset: Zoomed in
    plot for $R(r)$ near $r=0$.}
\end{figure}

For spirals, the topological singularity present at the core prevents us from obtaining the numerical values of $a_1$ and $b_1$. The small $r$ numerical profiles of $R(r)$ and $k(r)$ are plotted \cref{fig:smallr-spirals}(a). As the Cartesian grid used to simulate the defect solutions of the NRCH model lacks the polar symmetry of the defect at the origin, $R(r)$ never truly vanishes at the origin, and the change in polar angle $\Delta\theta=\pm\pi$ as one crosses the origin (for example on $y=0$ line) appears as a real phase change within a few units of grid spacing $\Delta x = L/N$ and leads to a numerically diverging $k(r)$ at $r=0$ for spirals. In \cref{fig:smallr-spirals}(b), we show that this is an unavoidable numerical artefact on Cartesian grids which reduces slowly as we improve the grid resolution. Note that, spirals for the CGLE equation have similar features. To get the correct numerical profiles for spirals, one can directly solve \eqref{eq:frgr} with appropriate boundary conditions as a nonlinear eigenvalue problem, we leave the problem open for future.

\begin{table}[b]
    \centering
    \begin{tabular}{@{\extracolsep{\fill}} c c c}
        \hline
        Box length ($L$) & & Spiral states\\
        \hline
                         & $\chi'=0$ & $\chi'=1/2$\\
                         \hline
        $100$ & $0\%$ & $0\%$\\
        $200$ & $15\%$ & $53\%$\\
        $400$ & $85\%$ & $88\%$\\
        $800$ & $100\%$ & $100\%$\\
        \hline
    \end{tabular}
    \caption{
        \label{tab:finite} Percentage of simulations of \eqref{eq:nrch-full} showing defect-laden steady states at different box sizes. Parameters: $a=1,~b=1,~\chi=0, ~\alpha=0.2, K=1$. For each parameter set, we run 32 independent realizations of random initial conditions.
    }
\end{table}

\FloatBarrier

\clearpage

\section{\label{sec:finite} Defect solutions for the general NRCH model and finite-size effects}
In this section we show that defect solutions are a general feature of the NRCH model and are not restricted to the minimal model studied in the main text. We also show the effect of box size on the evolution of a disordered state to various steady-states.

\begin{figure}[b]
    \centering{\includegraphics[width=0.9\linewidth]{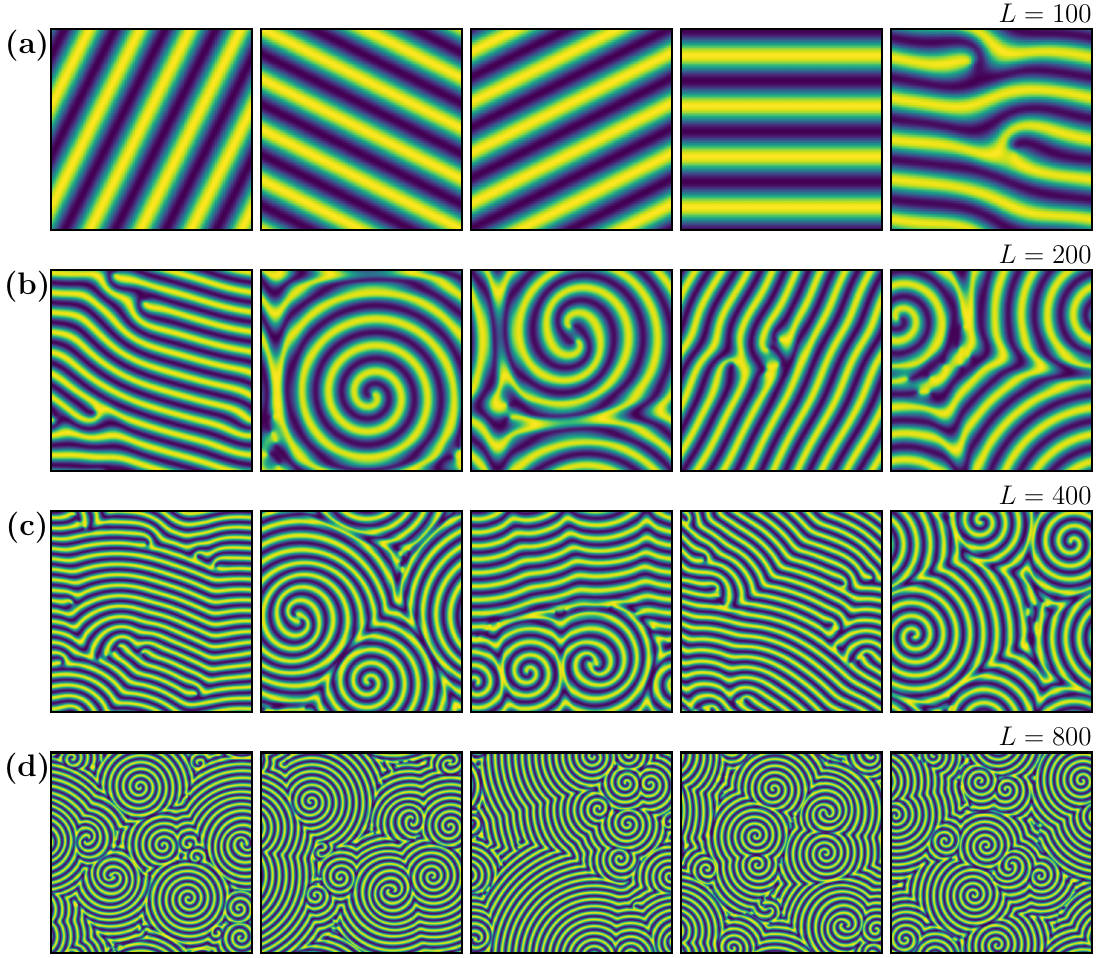}}
    \caption{
        \label{fig:boxsize_0.5} Representative snapshots in steady states for varying box sizes for $\chi=0$, $\chi' = 1/2$, and $\alpha = 0.2$. Each snapshot represents an independently evolving simulation initialized with different random initial condition. At small boxes, we only observe plane waves. With increasing $L$, we find spirals.
    }
\end{figure}

\begin{figure}[t]
    \centering{\includegraphics[width=0.9\linewidth]{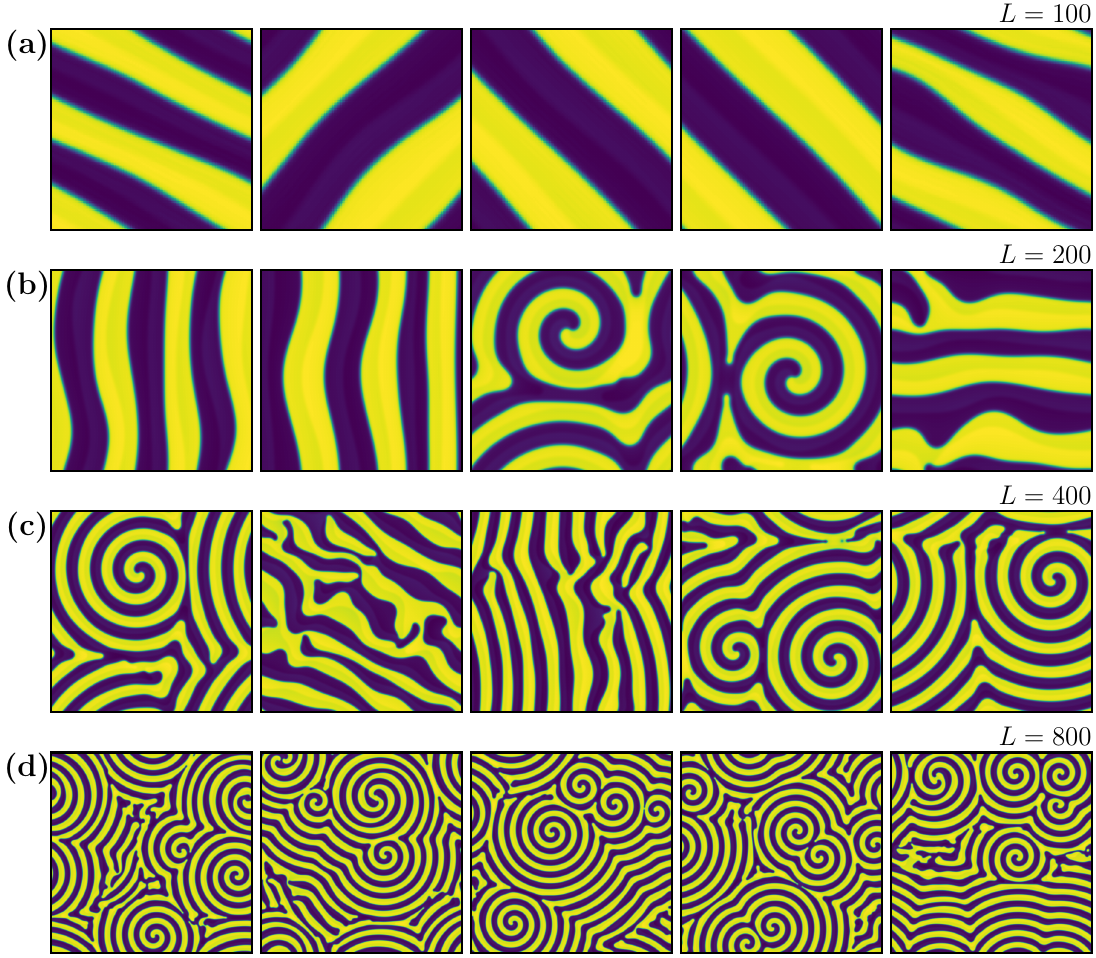}}
    \caption{
        \label{fig:boxsize_0.0} Representative snapshots in steady states for varying box sizes for $\chi=0$, $\chi' = 0$, and $\alpha = 0.2$. Similar to the $\chi' = 1/2$ case, finite size effects eliminate spirals for small box sizes.
    }
    \centering{\includegraphics[width=0.9\linewidth]{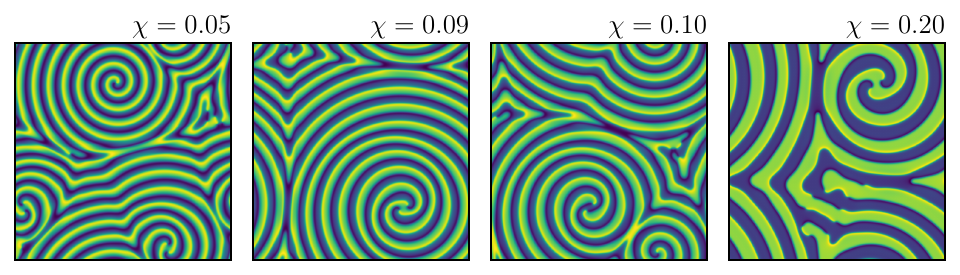}}
    \caption{ \label{fig:varying_chi}
        Spiral solutions for different values of $\chi$ for fixed $\alpha=0.1$ and $\chi' = 1/2$.
    }
\end{figure}

In \cref{fig:boxsize_0.5} and \cref{fig:boxsize_0.0}, we plot the steady states observed in an ensemble of simulations of \eqref{eq:nrch-full} for $\chi'=\frac{1}{2}$ and $\chi'= 0$ at various box sizes. For the smallest box $L = 100$ (\cref{fig:boxsize_0.5}(a) and \cref{fig:boxsize_0.0}(a)) the steady state for all the simulations is a travelling wave.
As we increase the box size, the probability of finding defect solutions in the steady state increases, and for $L=800$ (\cref{fig:boxsize_0.5}(d) and \cref{fig:boxsize_0.0}(d)) all the simulations show spiral-laden steady states. In \cref{tab:finite} we list the percentage of defect-laden steady states versus the box size $L$. In \cref{fig:varying_chi}, we show spiral solutions obtained when the interactions are not entirely anti-symmetric i.e., $\chi \neq 0$. Thus, we conclude that various parameters of \eqref{eq:nrch-full} and the box size determine the nature of steady states, but defects in general are the preferred solutions for large system size. Travelling waves are observed for smaller domains, as was the case in the previous work \cite{saha2020}.

\section{\label{sec:numerics} Numerical methods}

We use a GPU-accelerated pseudo-spectral algorithm to integrate the equations of motion \eqref{eq:nrch-full} or \eqref{eq:nrch-supp} numerically in a two-dimensional square periodic box of side $L$ discretized over $N^2$ collocation points. For all values of $L$ and $N$ reported, we fix $\Delta L = L/N = 1.5625$. For time marching, we use a second-order exponential time differencing scheme \cite{cox2002}. To study the properties of isolated defect solutions, we initialize our simulations with the ansatz \[\phi\left(\bm{r}, 0\right) = e^{i \left(m \theta - k r\right)},\] with the origin at center of the box. We initialize our quenched disordered simulations with random numbers generated from a uniform distribution with zero mean and $1/12$ variance. We further set $\left<\phi\right>=0$ at the begining of all our simulations. We have verified that the same asymptotic wavenumber $\kl$ selected in both simulations starting from random initial conditions and in isolated defect simulations irrespective of the initial value of $k$.

\section{\label{sec:movies} Description of the movies}
\begin{itemize}
    \item \texttt{\textbf{01\_isolated\_defects.mkv}} shows the evolution of an isolated target and spiral for $\alpha=0.15$. The target pulsates, and the spiral arm rotates. Both defects generate radially outward travelling waves.
    \item \texttt{\textbf{02\_defect\_networks.mkv}} shows the formation of a defect network at $\alpha=0.12$. We plot $\mathrm{Re}\left(\phi\right)$ to highlight the oscillatory nature of the defect network and $|\phi|$ to show the movement of defect cores. After an initial transient period, the defect network becomes quasi-stationary with dynamics limited to the disinclination lines.
    \item \texttt{\textbf{03\_unstable\_ring.mkv}} shows the evolution of target with initial wavenumber larger than the asymptotically selected wavenumber. Due to finite size effects and the interference of the emanated waves, the simulation evolves through a variety of intriguing patterns before eventually destabilizing.
\end{itemize}

\bibliography{Bibliography}
\bibliographystyle{apsrev4-2}